\PassOptionsToClass{10pt}{revtex4-2}
\documentclass[aps,prl,showpacs,floatfix,twocolumn,byrevtex,superscriptaddress]{revtex4-1}

\usepackage[utf8]{inputenc}
\usepackage{amsmath}
\usepackage{amssymb}
\usepackage{amstext}
\usepackage{amsopn}
\usepackage{amsfonts}
\usepackage{amsxtra}
\usepackage{fixltx2e}
\usepackage[english]{babel}
\usepackage{graphicx}
\usepackage{float}
\usepackage{bm}
\usepackage{multirow}
\usepackage{dcolumn}
\usepackage{color}
\usepackage{hyperref}
\usepackage{todonotes}
\usepackage{verbatim}
\usepackage{soul}
\usepackage{xcolor}

\def\FST{FeSe$_{1-x}$Te$_{x}$}

\date{\today}

\begin{document}
\author{D. Nevola}\email[]{nevola@bnl.gov}
\affiliation{Condensed Matter Physics and Materials Science Department, Brookhaven National Laboratory, Upton, New York 11973, USA}
\author{N. Zaki}
\affiliation{Condensed Matter Physics and Materials Science Department, Brookhaven National Laboratory, Upton, New York 11973, USA}
\author{J.M. Tranquada}
\affiliation{Condensed Matter Physics and Materials Science Department, Brookhaven National Laboratory, Upton, New York 11973, USA}
\author{W.-G. Yin}
\affiliation{Condensed Matter Physics and Materials Science Department, Brookhaven National Laboratory, Upton, New York 11973, USA}
\author{G.D. Gu}
\affiliation{Condensed Matter Physics and Materials Science Department, Brookhaven National Laboratory, Upton, New York 11973, USA}
\author{Q. Li}
\affiliation{Condensed Matter Physics and Materials Science Department, Brookhaven National Laboratory, Upton, New York 11973, USA} \affiliation{Department of Physics and Astronomy,
Stony Brook University, Stony Brook, New York 11794-3800, USA}

\author{P. D. Johnson}\email[]{pdj@bnl.gov}
\affiliation{Condensed Matter Physics and Materials Science Department, Brookhaven National Laboratory, Upton, New York 11973, USA}

\title{Ultrafast Melting of Superconductivity in an Iron-Based Superconductor}

\begin{abstract}
Intense debate has recently arisen regarding the photoinduced changes to the iron-chalcogenide superconductors, including the enhancement of superconductivity and a metastable state. Here, by employing high energy resolution, we directly observe the melting of superconductivity on ultrafast timescales. We demonstrate a distinctly nonequilibrium response on short timescales, where the gap fills in prior to the destruction of the superconducting peak, followed by a metastable response. We propose that the former is due to pair phase decoherence and speculate that the latter is due to the increase in double stripe correlations that are known to compete with superconductivity. Our results add to exciting new developments on the iron-based superconductors, indicating that the photoinduced metastable state possibly competes with superconductivity.
\end{abstract}
\maketitle

\section{Introduction}
The superconducting iron chalcogenides have recently generated considerable attention with the recognition that they bring together superconductivity, topology and magnetism in a single material\cite{Zhang2018,Rameau2019,Zaki2021}. Superconductivity and topology offer the possibility of topological superconductivity which in turn presents a platform for hosting Majorana Fermions, potentially important in qubit technology\cite{Wang2018}. Magnetism and topology on the other hand offer a platform for the Quantum Anomalous Hall Effect (QAHE). However, superconductivity and the QAHE represent competing states\cite{Wu2021}, and it is therefore important that an understanding is obtained of the different phenomena driving the system one way or the other. The \FST{} (FST) phase diagram is complex with the system displaying superconductivity for x = 0 and antiferromagnetism for x = 1. At around x = 0.5 and higher the possibility of a topological state existing on the surface has also been predicted\cite{xu2016}. The chiral nature of the latter topological surface state (TSS) has been demonstrated using spin polarized photoemission in one study\cite{Zhang2018} and incident circularly polarized light in another\cite{Rameau2019}. It has also been shown that for x $>$ 0.5 ferromagnetism develops in the surface region with the superconducting transition\cite{Nathan2021}. The associated time reversal symmetry breaking (TRSB) results in a gap at the Dirac point in the TSS\cite{Zaki2021}.

There have also been a number of studies of this material using time-resolved pump-probe techniques where the system is pumped out of equilibrium and the pathway back to equilibrium provides further insights into the complex interactions that take place between the electrons and collective modes\cite{Luo2012,Bonavolonta2013,yang2015,Gerber2017,Suzuki2019,Yang2021}. 

Such techniques also have been applied to the cuprate superconductors for Fermi surface wave vectors ranging from the nodal direction to the anti-nodal direction. In general, these studies show a return to equilibrium via different stages. First the non-equilibrium state decays via e-e interactions on time scales of the order of ~100 femtoseconds, establishing a state with a “pseudo-equilibrium" Fermi-Dirac distribution\cite{Perfetti2007,cortes2011,smallwood2012,Parham2017,piovera2015,rameau2014,konstantinova2018, rameau2016}. At longer time-scales, of the order of picoseconds, energy is subsequently transferred to the lattice via electron-phonon interactions. Studies have also concluded that with IR excitation, the superconducting gap fills in, rather than closing, due to loss of coherence of the superconducting condensate near the node\cite{Boschini2018,piovera2015}, while more exotic behavior due to Mott physics may be present at the antinode\cite{dakovski2015,Cilento2018}.

Pump-probe studies of the Fe-chalcogenide superconductors show a different behavior.  These systems still show the fast electron-electron equilibration but then the material appears to enter a metastable phase that exists on the order of nanoseconds or longer\cite{Fanfarillo2021,Yang2021,Suzuki2019}. The nature of this metastable behavior has been the topic of considerable discussion. Photoemission studies have speculated about various behaviors, ranging from light-induced superconductivity\cite{Suzuki2019} to light-induced nematicity\cite{Fanfarillo2021}. The concept of light-induced or enhanced superconductivity appears to follow the pioneering studies on this subject in a number of materials by Cavalleri and coworkers\cite{mankowsky2014,mitrano2016,mankowsky2017,budden2021}. By the very nature of the experimental probes, these studies have tended to focus on high timing resolution.  In the present study we take a different approach.  Acknowledging that the light induced state is metastable and therefore long lived, we \textit{reduce the timing resolution and push for high energy resolution in the probe beam}.  Such an approach offers immediate and important new insights into the metastable state. Not only can we now for the first time identify the coherent peak associated with the bulk superconductivity in a time-resolved study of this material, but we clearly demonstrate from its transient behavior that the incident infrared (IR) pulse kills the superconductivity and that it does not recover in the time scale associated with the metastable phase.  

In systems with closely competing quantum states, it is known that an incident IR pulse can push the system from one state to another.  In the FST materials, it is well established that two different spin or magnetic configurations exist.  The low temperature superconducting phase is characterized by spin correlations of the single-stripe type\cite{Li2021}. At slightly higher temperatures above 40~K the material adopts double-stripe spin correlations that do not support superconductivity\cite{tranquada2020}.  Equilibrium studies have suggested that this second phase is closely related to the electron coupling with the $A_{1g}$ phonon mode, which is the out-of-plane oscillation of the chalcogen atoms\cite{Liu2013}. Interestingly, the out-of-plane chalcogen position has been theoretically tied to the magnetic phase, supporting the significant role of the electron-$A_{1g}$ phonon coupling in determining the magnetic ground state\cite{yin2010,yin2012}. Similarly, out-of-equilibrium studies have shown that the pump-induced, metastable state is distinctly related to the $A_{1g}$ phonon mode, eventually leading to a renormalized chalcogen position\cite{Gerber2017,Yang2021}. This renormalized position shows an increased c-axis lattice constant, consistent with increasing the value of \textit{x} on the \FST{} phase diagram\cite{tranquada2020}. It is then naturally expected that this metastable state will lead to an associated modification of the superconducting and magnetic behavior. In the present study we  find that superconductivity is not supported in this metastable phase, and therefore postulate that it reflects a transition from the superconducting single-stripe phase to the double-stripe non-superconducting state, driven by photoinduced orbital fluctuations.
\section{Methods}
Single crystals were grown by the unidirectional solidification method. The superconducting properties of the single crystals studied here were previously characterized by magnetization measurements, having a superconducting transition temperature, T$_c$ of 14.5~K\cite{Zaki2021}.

The time-resolved ARPES was performed at Brookhaven National Laboratory using a SES-2002 analyzer. The probe pulse was generated using the output of a Coherent Vitara-T/RegA 9050 setup, which consists of 70~fs long, 800~nm pulses at 250~kHz. A cascade of nonlinear process was then used to produce the fourth harmonic (200~nm). The fourth harmonic pulses were finally passed through a monochromator before entering the vacuum chamber. The 800~nm fundamental pulses were used to pump the system. The energy and time resolutions were 6.6~meV and 500~fs, respectively. Space charge effects were carefully monitored before each experiment. The temperature-dependent measurements were acquired first at 5~K, then by increasing to 15~K, and finally reproduced again at 5~K on the sample spot, in order to ensure that the effect was reproducible over a full temperature cycle. The results in this paper have been repeated on multiple cleaves.

The fits in Fig.~\ref{Fig3}c consisted of the sum of two exponentially modified Gaussians (EMG's), which is the analytical form for an exponential convolved with a Gaussian. The decay time of the longer one was fixed to be 1~ns in order to mimic the constant offset, and the width of the Gaussian corresponds to the time resolution. The fits in Fig.~\ref{Fig3}d consisted of the sum of an error function and EMG, where the widths in both correspond to the resolution and were fixed to be the same.

\begin{figure}[tb]
\includegraphics[width=8.6 cm]{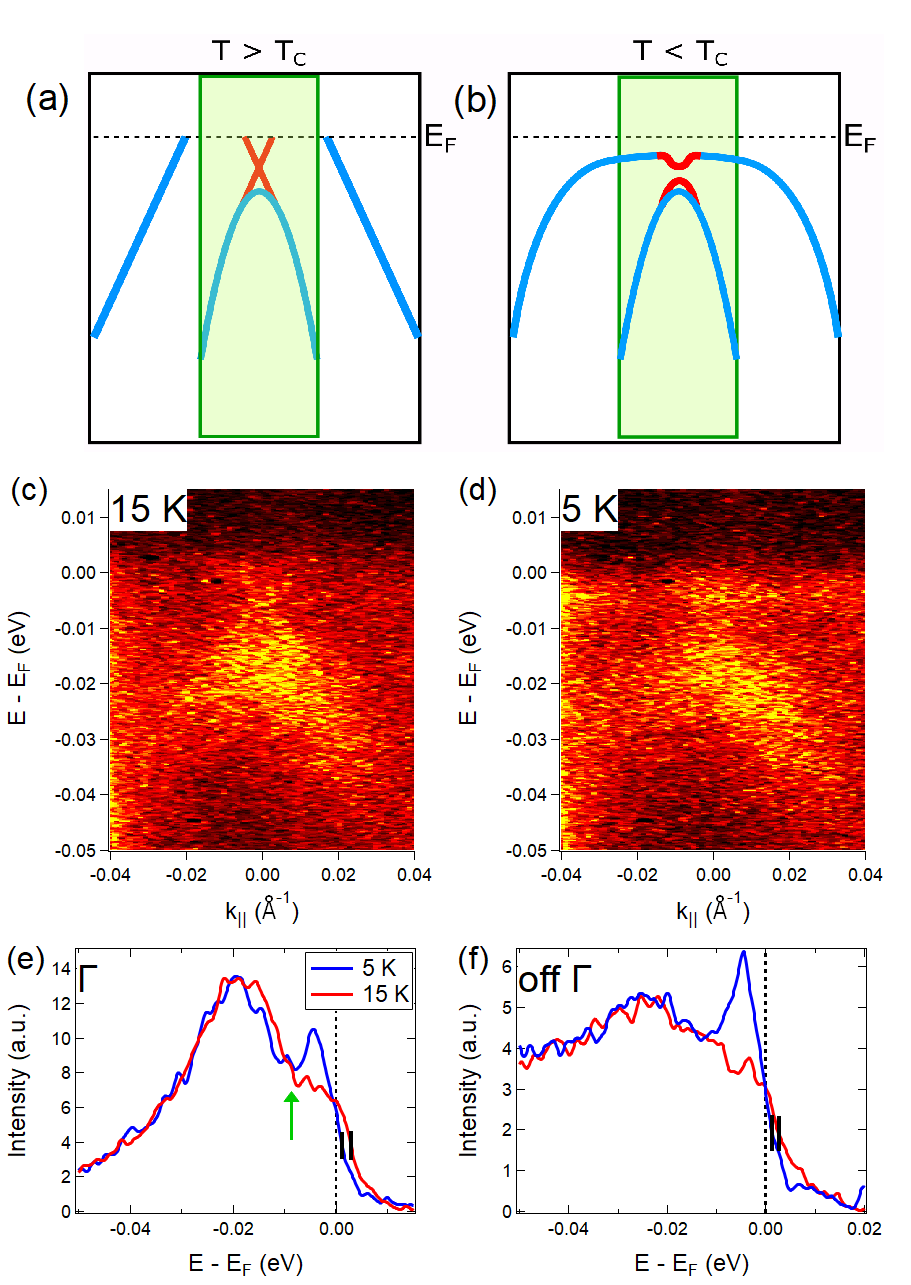}
\caption{Band structure of FeSe$_{0.45}$Te$_{0.55}$. (a) and (b) schematically show the band structures near the $\Gamma$ point above and below T$_c$, respectively, and are based off of the results in References~\cite{Zhang2018,Rameau2019,Zaki2021}}. The bands in blue and red represent bulk and surface contributions. The green box is representative of the angular window probed by the experiment. Static ARPES, probed with p-polarized light at 15~K (c) and 5~K (d). Energy distribution curves centered at $k=0$ (e) and $k=0.03~\r{A}$ (f). The momentum integration windows were $0.03\r{A}^{-1}$ and $0.02\r{A}^{-1}$ for (e) and (f). Red and blue EDC's represent those at 15~K and 5~K, respectively. The data in (e) and (f) has been smoothed for clarity.
\label{Fig1}
\end{figure}

\section{Band Structure}
We start by reviewing the changes to the band structure near the $\Gamma$ point though the superconducting transition, schematically shown in Fig.~\ref{Fig1}a,b (for a more comprehensive understanding of the band structure, the reader is directed to Refs.\cite{Zhang2018,Zaki2021,Rameau2019}, from which the schematics in Figures~\ref{Fig1}a,b are based). We focus mainly on the three bands closest to the zone center, which is comprised of two bulk bands and a topological surface state. Above T$_c$, the outer-most bulk band forms a hole pocket that crosses E$_F$ near $\pm0.15~\r{A}^{-1}$\cite{Johnson2015}. At the zone center is another hole-like band that reaches its maximum $\sim~16~$meV below E$_F$\cite{Zaki2021,Zhang2018,Rameau2019}. These bulk bands are typically referred to as the $\alpha_2$ and $\alpha_1$ bands, respectively. The topological surface state emerges from the top of the $\alpha_1$ band to form a small, electron-like pocket\cite{Zhang2018,Zaki2021,Rameau2019}. Recent efforts have demonstrated that the band inversion leading to the TSS results from the combination of spin-orbit effects in the Te atoms and hybridization from the Fe d$_{xz/yz}$ and Te p$_z$ orbitals\cite{lohani2020,xu2016}. Theoretical efforts to reproduce the ground state band structure highlight the role of strong electronic correlations that we believe play a vital role in the metastable behavior observed in time-angle resolved photoelectron spectroscopy (trARPES), as will be discussed in more detail later. For example, in one study, density functional theory requires a strong renormalization to reproduce the experimentally observed energy scales\cite{lohani2020}, while another study directly shows that the inclusion of magnetic correlations overcomes this discrepancy\cite{Johnson2015}. 

The system undergoes several changes upon cooling through the superconducting transition (Fig.~\ref{Fig1}b). First, superconducting gaps open up in both the bulk and surface Fermi pockets, with values of $2\Delta\sim$~4 and 3.6~meV respectively\cite{Okazaki2014,Zhang2018}. Additionally, a weakly dispersive superconducting peak (SP) develops in the vicinity of E$_F$\cite{Lubashevsky2012,Okazaki2014,Hashimoto2020}. At the same time, a 5~meV gap opens at the Dirac point in the TSS and has been used to associate the onset of superconductivity with the formation of ferromagnetic order\cite{Zaki2021}. The polarization-dependent matrix elements have proved a useful method for enhancing or suppressing these bands. For example, s-polarized light enhances the $\alpha_1$ band and the TSS, whereas p-polarized light enhances the $\alpha_2$ band and the SP\cite{Ronott2017,Zaki2021}. As we will argue later, such a strong polarization dependence of the SP may reflect its sensitivity to orbital fluctuations, which can be strongly altered by even a weak photoexcitation. For the present study, we probe with p-polarized light because we are interested in the pump-induced changes to superconductivity.

\section{Results}
Next, we demonstrate the high energy resolution of the trARPES experimental configuration by showing the static ARPES results above and below T$_c$ respectively  (Figs.~\ref{Fig1}c,d). The low energy scales associated with the bands of interest emphasize the requirement for such resolution. Above T$_c$, the $\alpha_1$ band is clearly visible. However, we note that the $\alpha_2$ band is outside of our momentum window, which is shown schematically by the green rectangles in Fig.~\ref{Fig1}a,b. We also observe the TSS as a weak intensity plateau extending from the top of the $\alpha_1$ band. This is more clearly seen in the energy distribution curve (EDC) through the $\Gamma$ point in Fig.~\ref{Fig1}e, where the green arrow indicates the Dirac point. Above the Dirac point at 15~K, the plateau corresponding to the TSS is visible. As alluded to earlier and discussed elsewhere, below T$_C$ two gaps appear in the spectral response, one at the chemical potential associated with superconductivity and one at the Dirac point reflecting TRSB. We are unable to clearly resolve the topological gap because a better momentum resolution is required than available in the present study. The superconducting gap is seen by observing the shift of the leading edge midpoints, shown by the black ticks in figure~\ref{Fig1}e,f, by 2.5~meV\cite{Zaki2021}. The formation of the superconducting gap results in the development of a peak just below the Fermi level. Finally, there is a subtle shift towards higher binding energy of the $\alpha_1$ band that occurs in conjunction with the formation of the gap at the Dirac point (Fig.~\ref{Fig1}e). All of these changes are also made evident when applying the Lucy-Richardson deconvolution algorithm \cite{Rameau2010}[see Supplemental]. Having established the ability to resolve the low-energy effects associated with superconductivity, we now turn to the pump-induced modifications to the superconducting band structure.

\begin{figure}[tb]
\includegraphics[width=8.6 cm]{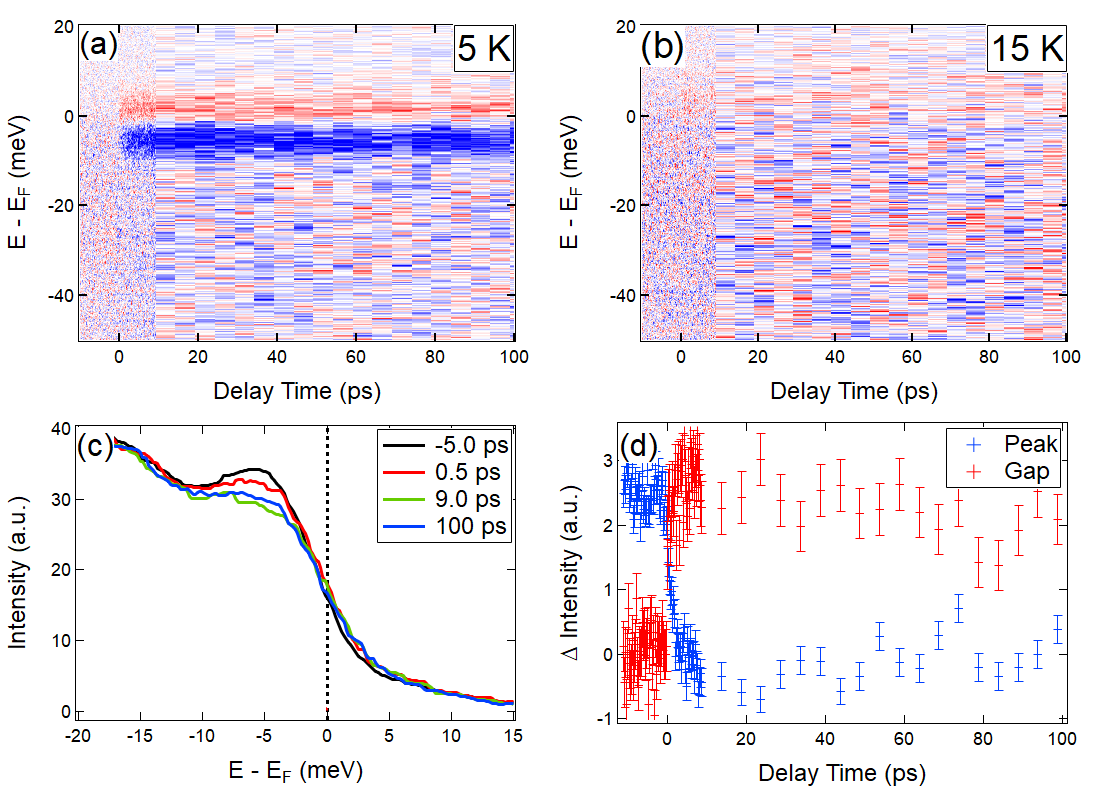}
\caption{Photoinduced changes to the k-integrated band structure below (a) and above (b) T$_c$ at an incident pump fluence of 1.2~$\mu$J/cm$^2$. Red and blue indicate an intensity increase and decrease respectively. (c) Energy distribution curves at various delay times. (d) Intensity dynamics just above and below E$_F$ out to long delay times. The data was averaged over a 7~meV energy range. For clarity, the data below E$_F$ was offset and normalized to match the change above E$_F$.}
\label{Fig2}
\end{figure}

Figure~\ref{Fig2}a shows the k-integrated changes to the band structure below T$_c$ upon pumping with an infrared laser pulse. Two features are observed that are absent when pumped above T$_c$ (Fig.\ref{Fig2}b): 1) an intensity increase just above E$_F$ and 2) a sharp intensity drop just below E$_F$. As these changes are not observed above the superconducting transition, they must be associated with superconductivity. Examination of the momentum-integrated EDC's in Fig.~\ref{Fig2}c shows that these intensity changes correspond to the filling of the superconducting gap, and destruction of the SP, respectively. These changes are consistent with the static results in Fig.~\ref{Fig1} and indicate that superconductivity is melted on ultrafast timescales. Identifying the mechanisms responsible for this phenomenon, as well as its recovery, is a novel pathway towards understanding iron-based superconductivity. We now turn to the time dependence for further insight into these processes.

Figure~\ref{Fig2}d shows that after the initial melting of superconductivity, the changes persist up to $>~100$~ps without showing any clear sign of recovery. However, we note that slight deviations from a constant offset after the first few picoseconds can be attributed to noise, slight drift of the pump beam, or acoustic oscillations that have been observed to occur on these timescales\cite{Luo2012}. This is also seen when comparing the k-integrated EDC's at 9~ps to that at 100~ps in Fig.~\ref{Fig2}c. Indeed, similar long time behavior has been observed in other optical experiments that have used similar pump fluences to the ones used in this work\cite{Luo2012}. This lack of decay on $\sim$100~ps timescales is intriguing when considering the metastable behavior that has attracted recent interest\cite{Fanfarillo2021,Yang2021,Suzuki2019}. Importantly, the main difference between those studies and the present one is the pump fluence used (much lower in the present case). Furthermore, the linear fluence dependence of the metastable changes with no threshold establishes this long-lived behavior as a derivative of the ground state, meaning that its existence owes to the interactions present in the ground state\cite{Gerber2017}. Thus, the decay bottleneck that is ultimately responsible for the metastable behavior must still be present at the low fluences used in this work. We believe that the reduction of superconductivity is a property of the pump-induced electronic redistribution, and the long-lived behavior results from the same mechanism as the well-reported metastability in FST when pumped at higher fluence\cite{Gerber2017,Fanfarillo2021,Suzuki2019,Yang2021}. However, it is also expected that pumping a superconductor at T~$<$~T$_c$ naturally leads to the melting of superconductivity on picosecond timescales due to equilibrium heating from the incident laser pulse. The dynamics that we observe at shorter delays do not show the expected picosecond recovery typically associated with lattice heating and must therefore correspond to time-dependent changes to the spectral function, $A(k,\omega,t)$.

\begin{figure}[tb]
\includegraphics[width=8.6 cm]{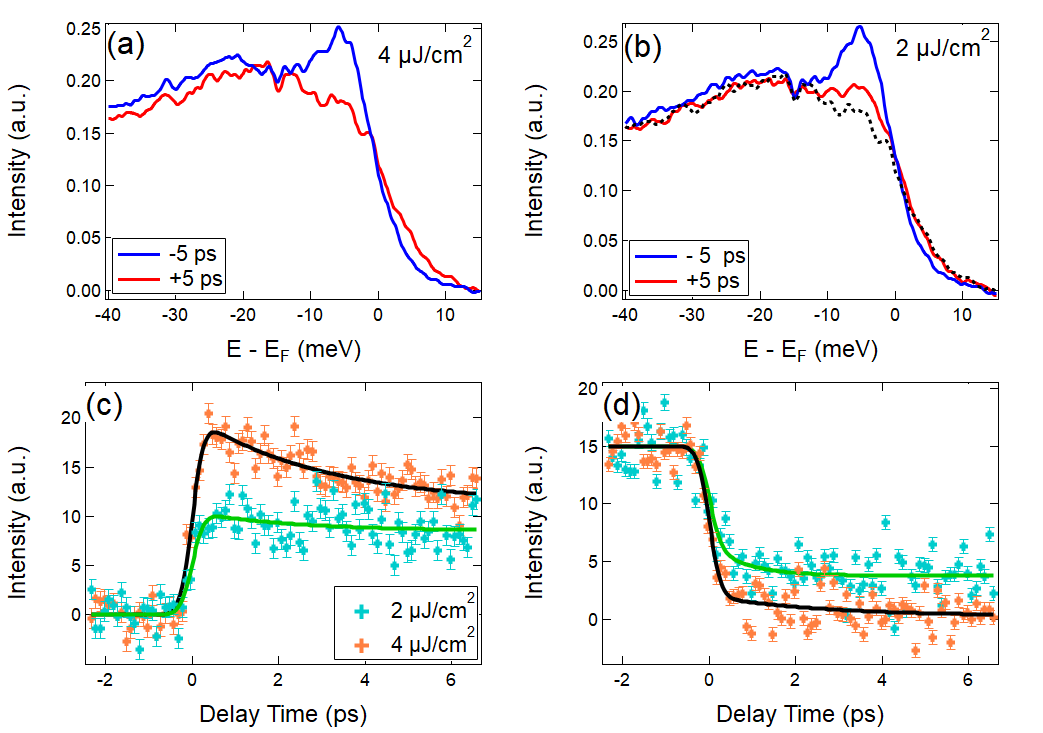}
\caption{Fluence-dependent dynamics at short delay times. (a) EDC's before (blue) and after (red) the pump pulse at 4~$\mu$J/cm$^2$. (b) Same as (a), but with a pump fluence of 2~$\mu$J/cm$^2$. The dashed black line is the red curve in (a) for comparison. Dynamics of the momentum integrated, difference images just above (c) and below (d) E$_F$ that correspond to the superconducting gap and quasiparticle peak, respectively. Solid lines correspond to fits.}
\label{Fig3}
\end{figure}

We consider the results for two different fluences. Figure~\ref{Fig3} shows the dynamics at shorter delays for two different pump fluences. At the higher fluence of 4~$\mu$J/cm$^2$, the superconducting peak is destroyed, as shown by the difference between the EDC’s at negative and positive delay times recorded away from the zone center (Fig,~\ref{Fig3}a). However, most of the reduction is achieved with the lower fluence of 2~$\mu$J/cm$^2$ (Fig.~\ref{Fig3}c), which is made clear when comparing the off $\Gamma$ EDC at +5~ps (red curve) to the same EDC at 4~$\mu$J/cm$^2$ (dashed black curve). Comparing the amplitudes at +5 ps for the two fluences allows us to estimate the complete disappearance of the SP at a critical fluence of 3~$\mu$J/cm$^2$ (assuming a linear dependence on fluence).

Figures~\ref{Fig3}c,d, show the intensity dynamics at the indicated fluences above and below E$_F$, respectively. Careful analysis of the dynamical data reveals rapid changes equal to the resolution, followed by fluence-dependent picosecond dynamics. The initial change in signal is attributed to a combination of the direct transitions associated with the pump, and subsequent electron-electron interactions that are expected to be reasonably strong in the iron-based superconductors. Although we are only able to determine an upper bound to this timescale to be 500~fs due to our resolution, reflectivity studies have shown that these initial changes occur within 100~fs\cite{Bonavolonta2013,Luo2012}. At the higher fluence, the positive signal (orange markers in Fig.~\ref{Fig3}c) shows a partial recovery on a timescale of 3~$\pm$~1.5~ps, while the negative signal (orange markers in Fig.~\ref{Fig3}d) shows an initial change equal to the resolution followed by no observable recovery. We interpret the 3~ps timescale as the dissipation of heat into the lattice. However, that this same timescale is not observed in the lower fluence means that this energy dissipation corresponds to excess heat after superconducting order has been melted. Similarly, temperature dependent reflectivity studies on FST show a similar picosecond timescale that disappears near T$_c$\cite{Luo2012,Bonavolonta2013}.

\begin{figure}[tb]
\includegraphics[width=8.6 cm]{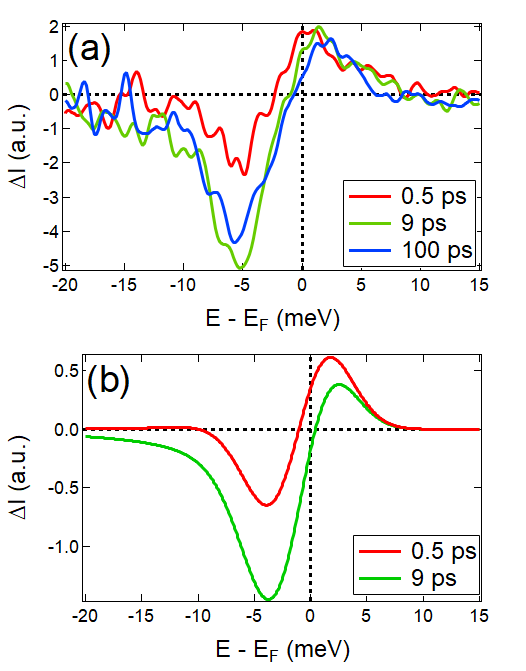}
\caption{(a)Time dependent energy profiles through the data in figure~\ref{Fig2}a at delay times of 0.5~ps, 9~ps and 100~ps. The pump fluence used was 1.2~$\mu$J/cm$^2$. (b) Modeled difference curves as described in the text. The red curve includes only a finite $\Gamma_p$ and the green curve includes both a finite $\Gamma_p$ and decreased spectral function amplitude, as further described in the supplementary information.}
\label{Fig4}
\end{figure}

The dynamics of the SP (Figure~\ref{Fig3}d) tell a slightly different story. Both fluences probed show no observed recovery on picosecond timescales. With the higher fluence, it would be expected that we would not observe the reappearance of the SP if there is excess energy after melting superconductivity. This can also be understood in the context of the two-temperature model with a final lattice temperature greater than T$_c$\cite{Allen1987}. If the melting of superconductivity was thermal, we would observe a recovery of the SP on a $\sim3$~ps timescale, similar to that above. Furthermore, it would be obvious since the SP is the brightest feature in the difference spectrum in figure \ref{Fig2}a, and is particularly sensitive to the temperature in equilibrium. Bonavolanta \textit{et al.} also determined the temperature to only increase by a maximum of 3~K when using a fluence several times those used in this work, making a thermal heating explanation unlikely\cite{Bonavolonta2013}. Closer examination of the lower fluence data reveals an initial decrease equal to the resolution followed by a further decrease on a timescale of 0.8~$\pm$~0.5~ps. This second timescale suggests that another scattering mechanism is involved in the melting of superconductivity as will be discussed in the next section. Additionally, since we remain below T$_c$ and observe no recovery of the condensate within the 100~ps measurement window, we conclude that the melting of superconductivity is metastable. 

To further emphasize the behavior at short delays, we display the difference EDCs (figure~\ref{Fig4}a), which are energy profiles through the data in figure~\ref{Fig2}a. For the intensity loss, we see an evolution in going from the shortest delay at 0.5~ps to 9~ps. However, for the intensity gain, we do not see the same evolution. After the maximum intensity change is reached within the first 0.5~ps, it remains constant to within the noise. This means that at short delays, the filling of the gap occurs first on timescales of $<~500$~fs.

\section{Discussion}
The most obvious implication of the results in this work is to demonstrate that superconductivity is melted on ultrafast timescales and does not return within the measured time window. This direct demonstration contradicts the conclusions in two recent works that have suggested enhanced superconductivity due to photoexcitation\cite{isoyama2021,Suzuki2019}. One is an all-optical study that suggests a temporary, picosecond, enhancement\cite{isoyama2021}, and another is a trARPES study with lower energy resolution that associates superconductivity with the metastable state\cite{Suzuki2019}. Interestingly, the former study does clearly demonstrate nonequilibrium behavior at short delays. 

Isoyama \textit{et al.} found an increase in both the real and imaginary parts to the conductivity within the first 1-2~ps, contradicting equilibrium measurements that showed an increase and decrease in the real and imaginary parts to the optical conductivity, respectively, with increasing temperature\cite{isoyama2021}. Additionally, these changes occur on an energy scale $<2$~meV, less than the superconducting gap, $2\Delta$, and the same magnitude of the observed leading edge midpoint shift(see Supplemental). One possible interpretation is an increase in the size of the condensate, as they suggested, but this would manifest as an increased spectral weight within the superconducting peaks on either side of the Fermi level, which we do not observe. Alternatively, we propose the destruction of phase coherence among the Cooper pairs caused by their scattering with nonequilibrium bosonic excitations\cite{Boschini2018}. This mechanism may explain both the increases to the real\cite{xi2010} and imaginary\cite{lemonik2018} parts of the optical conductivity. It can be modeled by the inclusion of a finite pair scattering rate in the self energy that describes a superconductor\cite{kwon1999,franz1998,Boschini2018}:

\begin{equation}
  \Sigma(\omega)=-i\Gamma_s+\frac{\Delta^2}{\omega+i\Gamma_p}.
  \label{eq1}
\end{equation}

Here, $\omega$ is the energy,  $\Delta$ is the gap amplitude, $\Gamma_s$ is the single particle scattering rate, and $\Gamma_p$ is the pair scattering rate, indicative of pair scattering with phase fluctuations\cite{Boschini2018}. An increase in $\Gamma_p$ increases the spectral weight inside the gap, and was recently found to describe the nonequilibrium changes to the near-nodal spectral function in an optimally doped cuprate\cite{Boschini2018}.

To provide a more quantitative picture, we modeled the difference curves by using the self-energy in equation~\ref{eq1} (see Supplemental for more details). First, we show the experimental data at three delay times in figure~\ref{Fig4}a (the data correspond to energy profiles at the specified delay times selected from the data in figure~\ref{Fig2}a). The amplitudes of the positive and negative signals are nearly identical in the 0.5~ps (resolution limited) data, while the negative signal grows on a picosecond timescale, and remains constant for $>$~100~ps. As shown in figure~\ref{Fig4}b, including a finite $\Gamma_p$, while keeping all other parameters constant, is enough to reproduce the experimental data at 0.5~ps. The data at longer delays was reproduced by reducing the amplitude of the spectral function, which we think resembles the breaking of Cooper pairs. That we are able to reproduce the curve shape at 0.5~ps by including the pair scattering rate, which is a nonequilibrium parameter, supports our hypothesis that the pump induces phase fluctuations caused by electronic correlations.

The photoinduced changes that we observed contain three timescales that offer further insight into the microscopic mechanisms responsible for the ultrafast melting of superconductivity. First, the gap filling occurs within our resolution. This suggests that the phase fluctuations are of electronic origin. The fluctuations may be caused by the scattering of Cooper pairs with nonequilibrium bosons\cite{Boschini2018}. We speculate that these bosons are of the form of magnetic excitations. Second, the persistent decrease in the SP occurs on a picosecond timescale, We attribute this to the breaking of pairs by optical phonons that are generated upon the decay of high energy quasiparticles. A similar, temperature dependent, timescale was observed in time-resolved reflectivity\cite{Bonavolonta2013} and was also ascribed to phonons. The final timescale is the recovery of the superconducting features, which are longer than our temporal window, indicating that the changes mentioned above are metastable. The key question is \textit{why} they are metastable, for which we offer two possible explanations that are based off of our observed timescales.

It is reasonable to assume that the mechanisms responsible for melting superconductivity also play a role in its recovery. As mentioned above, we observe electronic and phononic timescales that contribute to the melting of superconductivity.

The first explanation is that pairs are reformed by the absorption of low energy (acoustic) phonons, which can occur on nanosecond timescales. Indeed, this was used to explain the recovery of superconductivity after photoexcitation in the BCS superconductor MgB$_2$, which has similar gap sizes to the present material\cite{demsar2003}. However, two main issues with this explanation are that (1) even in the case of MgB$_2$, some recombination is observed on a 100~ps timescale, and (2) it neglects the role of electronic and magnetic correlations that are known to play a key role in mediating iron-based superconductivity\cite{tranquada2020}.

The second possibility is that the pairs are prevented from recombining because of the disruption to the correlations caused by the electronic redistribution. For example, direct excitations can occur at different locations throughout the Brillouin zone. The natural pathways for relaxation may cause most holes to relax to the $\Gamma$ point, and most electrons to the M point in the Brillouin zone. This redistribution of particles modifies the correlations that are key to mediating iron-based superconductivity. Though it is difficult to narrow down the specific correlations that would be impacted, examining the changes that occur at higher excitation densities, where the changes are more apparent, may offer further insight. Thus, we turn to recent works that have identified a photoinduced metastable state in FST for clues. Properties of this metastable state include the renormalizations of the band structure and chalcogen positions that resemble the ground state effects of increasing the Te concentration\cite{Gerber2017,Fanfarillo2021,Yang2021}.  We therefore hypothesize that the photoinduced changes to the correlations are similar to those in higher Te concentrations.

At higher Te concentrations, the correlations favor double stripe magnetism. Additionally, these magnetic correlations are known to compete with superconductivity\cite{tranquada2020}. We therefore hypothesize that the photoexcitation causes an orbital redistribution that favors double stripe magnetic correlations, which act to melt superconductivity. Indeed, the sensitivity of superconductivity to the orbital redistribution is supported by electron diffraction work on Ba(Fe$_{1-x}$Co$_x$)$_2$As$_2$ that directly images the orbital configuration\cite{ma2014}, as well as the strong polarization-dependent matrix elements of the SP observed in ARPES\cite{Rameau2019,Zaki2021} and scanning tunneling microscopy\cite{singh2015} because they demonstrate the strong orbital dependence of the superconducting bands. Additionally, theoretical studies have demonstrated how the subtle balance among the band filling, the nearest and next-nearest superexchange interactions, and the Hund’s coupling strength, which may be particularly sensitive to a pump-induced orbital redistribution, determine the magnetic ground state\cite{yin2010,yin2012}.

\section{Conclusion}
In conclusion, we directly show, for the first time, the pump-induced changes to the electronic band structure of FST below T$_c$. Our observations show a reduction of two features that are directly associated with superconductivity, namely the superconducting peak and gap. We observe changes both on electronic and picosecond timescales, and find that they are metastable, lasting for $>$~100~ps. Close examination of the initial changes to the band structure are indicative of nonequilibrium behavior that shows the filling of the superconducting gap occurring before the reduction of the SP. We hypothesize that the orbital fluctuations directly induced by the pump-induced electronic redistribution are linked to double stripe magnetic correlations known to  compete with superconductivity. A better theoretical understanding of the precise electronic excitations and relaxation pathways would be crucial in understanding the changes to the orbital occupancy that results from the pump. Furthermore, experiments aimed at directly imaging the photoinduced orbital and magnetic changes, as well as trARPES with high energy resolution extending out to the edge of the Brillouin zone, would be needed in order to corroborate this picture, and may ultimately yield clues to understanding the interactions that lead to iron-based superconductivity.

\section{Acknowledgements}
We acknowledge Chris Homes, Mengkun Liu, and Abhay Narayan Pasupathy for insightful discussions. Work at Brookhaven National Laboratory was supported by the U.S. Department of Energy, Office of Science, Office of Basic Energy Sciences, under Contract No. DE-
SC0012704.
\bibliography{ref}

\end{document}